\documentclass[
orivec,oribibl,envcountsect]{llncs}

\usepackage{amssymb}
\usepackage{graphicx}
\usepackage{mathrsfs}
\usepackage{stmaryrd}

\usepackage{amsmath}
\DeclareMathAlphabet{\mathpzc}{T1}{pzc}{m}{it}

\usepackage[all]{xy}
\UseComputerModernTips

\spnewtheorem*{remark*}{Remark}{\itshape}{\rmfamily}
\spnewtheorem*{notation}{Notation}{\itshape}{\rmfamily}

\newenvironment{mydescription}
{\begin{list}{}{\setlength{\leftmargin}{1.5em}\setlength{\labelwidth}{1em}}}{\end{list}}

\newcommand{\hide}[1]{}

\newcommand{\eg}{\emph{e.g.}}

\newcommand{\etc}{\emph{etc.}}
\newcommand{\ie}{\emph{i.e.}}

\newcommand{\eqdef}{=}
\newcommand{\Sig}{\Sigma}

\newcommand{\Oper}{O}
\newcommand{\oper}[1]{\mathsf{#1}}

\newcommand{\arity}[1]{\mid\!{#1}\!\mid}

\newcommand{\seq}[1]{\langle{#1}\rangle}
\newcommand{\bigseq}[1]{\big\langle{#1}\big\rangle}
\newcommand{\bind}[1]{({#1})}
\newcommand{\pair}[1]{\langle{#1}\rangle}
\newcommand{\bigpair}[1]{\big\langle{#1}\big\rangle}

\newcommand{\var}[1]{#1}
\newcommand{\metavar}[1]{\mbox{\sc #1}}
\newcommand{\sep}{\vartriangleright}

\newcommand{\card}[1]
{\lVert #1 \rVert}

\newcommand{\setof}[1]{\{{#1}\}}

\newcommand{\lscatfont}[1]{\mbox{\boldmath${#1}$}}
\newcommand{\scatfont}[1]{\mathbf{#1}}
\newcommand{\F}{\scatfont{F}}
\newcommand{\Set}{{\lscatfont{\mathcal{S}\hspace{-.5mm}\mathit{et}}}}
\newcommand{\SOAT}{{\lscatfont{\mathcal{SO}\hspace{-.5mm}\mathcal{AT}}}}
\newcommand{\SOEP}{{\lscatfont{\mathcal{SOE}\hspace{-.5mm}\mathcal{P}}}}
\newcommand{\yon}{\boldsymbol{y}}
\newcommand{\syon}{\yon}

\newcommand{\scat}[1]{\mathbb{#1}}

\newcommand{\comp}{\circ}
\newcommand{\icomp}{\,}
\newcommand{\id}{\mathrm{id}}
\newcommand{\lscat}[1]{\mathscr{#1}}
\newcommand{\op}{\mathrm{op}}

\newcommand{\stensor}{\bullet}

\newcommand{\coend}{\int}
\newcommand{\iso}{\cong}

\newcommand{\dom}{\mathrm{dom}}
\newcommand{\Con}{\Gamma}
\newcommand{\aCon}{\Delta}
\newcommand{\Mcon}{\Theta}

\newcommand{\msubst}[1]{\{ #1 \}}
\newcommand{\vsubst}[1]{[ #1 ]}
\newcommand{\bigvsubst}[1]{\big[ #1 \big]}
\newcommand{\subst}[2]{\raisebox{.75mm}{\small$#1$}\!/\!\mbox{\small$#2$}}

\newcommand{\lsem}{{[\![}}
\newcommand{\rsem}{{]\!]}}

\newcommand{\sem}[1]{\lsem{#1}\rsem}

\newcommand{\ul}{\underline}

\newcommand{\Mod}[1]{{#1}\mbox{-}\mathsf{Mod}}

\newcommand{\opclone}[1]{\langle #1 \rangle}

\newcommand{\Nat}{\mathbb{N}}

\newcommand{\EqTh}{\scatfont{M}}

\newcommand{\tup}[1]{({#1})}

\newcommand{\Exp}{\ShortRightarrow}

\newcommand{\FreeCart}{\scatfont{L}}

\newcommand{\eqpres}[1]{\mathcal{#1}}
\newcommand{\IntEqPres}{\mbox{\large$\mathpzc{E}$}}

\newcommand{\algth}[1]{\scat{#1}}

\newcommand{\Kl}{{\lscatfont{\mathcal{K}}}}

\newcommand{\eqclass}[1]{[ #1 ]}

\newcommand{\EndoFun}[1]{\mbox{\large$\mathpzc{F}$}\hspace{-1mm}_{#1}}
\newcommand{\sEndoFun}[1]{\mbox{\small$\mathpzc{F}$}\hspace{-1mm}_{#1}}
\newcommand{\TermMonad}[1]{\mbox{\large$\mathpzc{T}$}\hspace{-.75mm}_{#1}}

\newcommand{\trans}{\tau}

\newcommand{\FunMod}[2]
{{\lscatfont{\mathcal{M}\hspace{-.5mm}\mathit{od}}
(#1)}}

\newcommand{\wt}{\widetilde}

\newcommand{\shortarrowlength}{3} 
\newcommand{\arrowlength}{4}
\newcommand{\medarrowlength}{7}

\newcommand{\myarrow}[3] 
   {\mathrel{\xy\ar(#1,0)^-{#2}_-{#3}\endxy}}
\renewcommand{\rightarrow}
   {\myarrow{\arrowlength}{}{}}

\renewcommand{\leftarrow}
   {\myarrow{-\arrowlength}{}{}}

\newcommand{\stackrightarrow}[1]
   {\myarrow{\arrowlength}{#1}{}}
\newcommand{\midstackrightarrow}[1]
   {\mathrel{\xy\ar(\medarrowlength,0)|(.5125){#1}\endxy}}

\newcommand{\medstackrightarrow}[1]
   {\myarrow{\medarrowlength}{#1}{}}

\newcommand{\midstackleftarrow}[1]
   {\mathrel{\xy\ar(-\medarrowlength,0)|(.4){#1}\endxy}}

\newcommand{\mymidarrow}[2] 
   {\mathrel{\xy\ar(#1,0)|(.4){#2}\endxy}}

\newcommand{\rightembedding}
   {\mathrel{\,\xy\ar@{^(->}(\arrowlength,0)\endxy}}
\newcommand{\shortrightembedding}
   {\mathrel{\,\xy\ar@{^(->}(\shortarrowlength,0)\endxy}}

\newcommand{\myArrow}[3] 
   {\mathrel{\xy\ar@{=>}(#1,0)^-{#2}_-{#3}\endxy}}
\renewcommand{\Rightarrow}
   {\myArrow{\arrowlength}{}{}}

\newcommand{\ShortRightarrow}
   {\myArrow{\shortarrowlength}{}{}}

\newcommand{\relrightarrow} 
   {\mathrel{\xy\ar@{-|}(2.5,0)\endxy\xy\ar(3.5,0)\endxy}}

\newcommand{\mymapsto}[3] 
   {\mathrel{\xy\ar@{|->}(#1,0)^-{#2}_-{#3}\endxy}}
\renewcommand{\mapsto}
   {\mymapsto{\arrowlength}{}{}}

\newcommand{\shortrightmono}
   {\mathrel{\xy\ar@{>->}(\shortarrowlength,0)\endxy}}

\newcommand{\rightepi}
{\mathrel{\xy\ar@{->>}(\arrowlength,0)\endxy}}

\newcommand{\myleftrightarrow}[3] 
  {\mathrel{\xy\ar@{<->}(#1,0)^-{#2}_-{#3}\endxy}}
\newcommand{\shortleftrightarrow}
   {\mathrel{\xy\ar@{<->}(\shortarrowlength,0)\endxy}}

\newcommand{\noemph}[1]{#1}
\newcommand{\nombox}[1]{#1}

\newcommand{\shorten}[1]{}

\begin{document}

\frontmatter 

\title{Second-Order Algebraic Theories\\
{\large\text{\rm(Extended Abstract)}}
\\[-3mm]}

\author{Marcelo Fiore \and Ola Mahmoud}

\institute{University of Cambridge, Computer Laboratory
}

\maketitle

\mbox{}\\[-8mm]
%
\begin{abstract}
Fiore and Hur~\cite{FioreHur10} recently introduced a conservative extension
of \noemph{universal algebra} and \noemph{equational logic} from first to
second order.  \emph{\mbox{Second-order} universal algebra} and
\emph{\mbox{second-order} equational logic} respectively provide a model
theory and a formal deductive system for languages with variable binding and
parameterised metavariables.  This work completes the foundations of the
subject from the viewpoint of categorical algebra.  Specifically, the
paper introduces the notion of \emph{\mbox{second-order} algebraic theory} and
develops its basic theory.  Two categorical equivalences are established:
at the syntactic level, that of \mbox{second-order} equational
presentations and \mbox{second-order} algebraic theories; at the semantic
level, that of \mbox{second-order} algebras and \mbox{second-order}
functorial models.  Our development includes a mathematical definition of
syntactic translation between \mbox{second-order} equational
presentations.  This gives the first formalisation of notions such as
encodings and transforms in the context of languages with variable
binding.
\end{abstract}

\vspace*{-5mm}
\section{Introduction}

Algebra started with the study of a few sample algebraic structures:
groups, rings, lattices,~{\etc}  Based on these, Birkhoff~\cite{Birkhoff}
laid out the foundations of a general unifying theory, now known as
universal algebra. 

\hide{
Birkhoff's formalisation of the notion of algebra starts with the
introduction of the concept of an algebraic operator, a set of which forms
a signature.  Operators compose to give rise to terms~(or derived
operators), a pair of which constitutes an identity~(or axiom).  An
equational presentation for an algebraic structure is then a signature
together with a set of identities.  These constitute the syntactic
foundations of the subject.  On top of them, Birkhoff introduced algebras
as the semantics~(or model theory) of equational presentations.  An
algebra was thus defined as set equipped with interpretations for the
operators such that the induced interpretation of derived terms satisfies
the required identities.  
%
%
The model theory was then used by Birkhoff to establish the logical
foundations of the subject.  Indeed, by means of a completeness theorem,
Birkhoff introduced equational logic as the formal deductive system for
reasoning about algebraic structure.
}

Birkhoff's formalisation of the notion of algebra starts with the introduction
of equational presentations.  These constitute the syntactic foundations of
the subject.  Algebras are then the semantics or model theory, and play a
crucial role in establishing the logical foundations.  Indeed, Birkhoff
introduced equational logic as a sound and complete formal deductive system
for reasoning about algebraic structure.

The investigation 
of algebraic structure was further enriched by the advent of category
theory, with the fundamental work of Lawvere on algebraic
theories~\cite{LawvereThesis}
and of Linton on finitary monads~\cite{Linton}.
These approaches give a presentation-independent treatment of the subject.
Algebraic theories correspond to the syntactic line of development; monads
to the semantic one (see~\eg~\cite{HylandPower}).  

We contend that it is only by looking at algebraic structure from all of
the above perspectives, and the ways in which they interact, that the
subject is properly understood.  
In the context of computer science, for instance, consider that:
$(i)$~initial-algebra semantics provides canonical compositional
interpretations~\cite{ADJ};
$(ii)$~free constructions amount to abstract syntax~\cite{McCarthy}, that
is amenable to proofs by structural induction and definitions by
structural recursion~\cite{Burstall};
$(iii)$~equational presentations can be regarded as (bidirectional)
rewriting theories, and studied from a computational point of
view~\cite{KnuthBendix};
$(iv)$~algebraic theories come with an associated notion of algebraic
translation~\cite{LawvereThesis}, whose syntactic counterpart provides the
right notion of syntactic translation between equational
presentations~\cite{FujiwaraI,FujiwaraII};
$(v)$~strong monads have an associated metalogic from which equational
logics can be synthesised~\cite{FioreHurMFPS,FioreHur10}.

The realm of universal algebra is restricted to \mbox{first-order}
languages.  In particular, this leaves out languages with variable
binding.  Variable-binding constructs are at the core of fundamental
calculi and theories in computer science and
logic~\cite{ChurchLambda,ChurchTypes}, and incorporating them into algebra
has been a main foundational research problem.  The present work develops
such a programme from the viewpoint of algebraic theories.

Our presentation is in two parts.  The first
part~(Sections~\ref{SO:EquationalLogic} and~\ref{SO:UniversalAlgebra}) sets up
the necessary background; the second
part~(Sections~\ref{SO:AlgebraicTheories} to~\ref{SO:FunctorialSemantics})
constitutes the contribution of the paper.

The background material gives an introduction to the work of Fiore and
Hur~\cite{FioreHur10} on a conservative extension of universal algebra and
its equational logic from first to second order,~\ie~to languages with
variable binding and parameterised metavariables.  
Our summary recalls:
$(i)$~the notion of \mbox{second-order} equational presentation, that
allows the specification of equational theories by means of schematic
identities over signatures of variable-binding operators;
$(ii)$~the model theory of \mbox{second-order} equational presentations by
means of \mbox{second-order} algebras; and
$(iii)$~the deductive system underlying formal reasoning about
\mbox{second-order} algebraic structure.

The crux of our work is the notion of 
\emph{\mbox{second-order} algebraic
theory}~(Definition~\ref{SO:AlgebraicTheoriesDefinition}).  
%
%
At the syntactic level, the correctness of our definition is established
by showing a categorical equivalence between \mbox{second-order}
equational presentations and \mbox{second-order} algebraic
theories~(Theorem~\ref{SOAT-SOEP-Equivalence}).  This involves distilling
a notion of syntactic translation between \mbox{second-order} equational
presentations that corresponds to the canonical notion of morphism between
\mbox{second-order} algebraic theories.  These syntactic translations 
provide a mathematical formalisation of notions such as encodings and
transforms.
On top of the syntactic correspondence, we furthermore establish a semantic
one, by which \emph{\mbox{second-order} functorial semantics} is shown to
correspond to the model theory of \mbox{second-order} universal
algebra~(Theorem~\ref{Fun=Alg_Theorem} and Corollary~\ref{Fun=Alg_Corollary}).



\vspace*{-1.5mm}
\section{Second-Order Equational Logic}
\label{SO:EquationalLogic}

We briefly present \emph{\mbox{Second-Order} Equational Logic} as introduced
by Fiore and Hur~\cite{FioreHur10} together with the syntactic machinery that
surrounds it.  For succinctness, our exposition is restricted to the unityped
setting.  The general \mbox{multi-typed} framework can be found
in~\cite{FioreHur10}.

\vspace*{-4mm}
\subsubsection{Signatures.}
A (unityped \mbox{second-order}) \emph{signature} ${\Sig=(\Oper,\arity{-})}$
is specified by a set of operators~$\Oper$ and an arity function
${{\arity{-}}:\Oper\rightarrow\Nat^*}$, see~\cite{Aczel,AczelFS}.
For 
$\oper o\in\Oper$, we write $\oper o: \tup{n_1,\ldots,n_k}$ whenever
${{\arity{\oper o}} \,= \tup{n_1, \ldots, n_k}}$. 
The intended meaning is that the operator~$\oper o$ takes $k$ arguments with
the $i^\mathrm{th}$ argument binding $n_i$ variables.

\begin{example}
\label{LambdaCalculusSignature}
\ 
The signature of the 
\emph{\mbox{$\lambda$-calculus}} has operators ${\oper{abs}:(1)}$
and
\linebreak
${\oper{app}:(0,0)}$.
\end{example}

\vspace*{-5mm}
\subsubsection{Terms.} 
We consider terms in contexts with two zones, respectively declaring
metavariables and variables.  Metavariables come with an associated
natural number 
\noemph{arity}.  A metavariable $\metavar m$ of 
arity~$m$, 
denoted ${\metavar m:[m]}$, is to be parameterised by $m$~terms.
We 
represent contexts as
${\metavar m_1:[m_1] ,\ldots, \metavar m_k:[m_k] \sep x_1,\ldots,x_n}$ 
where 
the metavariables~$\metavar m_i$ and the variables~$x_j$
are assumed distinct.

Signatures give rise to terms in context.  Terms are built up by means of
operators from both variables and metavariables, and hence referred to as
\mbox{second-order}. 
The judgement for \emph{terms} in context ${(\Mcon\sep\Con\vdash -)}$ is
defined by the following 
rules. 
\begin{mydescription}
  \item[(Variables)]
    For $x\in\Con$,\\[-5mm]
    $$
    \begin{array}{c}
    \\ \hline
    \raisebox{-1mm}{$\Mcon\sep\Con\vdash \var x$}
    \end{array}
    $$

  \item[(Metavariables)]
    For $(\metavar m:[m])\in\Mcon$,
\vspace*{-1mm}
    $$
    \begin{array}{c}
      \raisebox{1mm}{$\Mcon\sep\Con\vdash t_i \enspace (1\leq i\leq m)$}
      \\ \hline
      \raisebox{-1mm}{$\Mcon\sep\Con
        \vdash \metavar m[t_1,\ldots,t_m]$}
    \end{array}
    $$
\vspace*{-3mm}

  \item[(Operators)]
    For ${\oper o: (n_1,\ldots,n_k)}$,
\vspace*{-1mm}
    $$
    \begin{array}{c}
      \raisebox{1mm}{$
      \Mcon\sep\Con,\vec{x_i}
        \vdash t_i
        \ (1\leq i\leq k)$}
      \\ \hline
      \raisebox{-1mm}{$\Mcon\sep\Con\vdash\oper
      o\big((\vec{x_1})\,t_1,\ldots,(\vec{x_k})\,t_k\big)
      $}
    \end{array}
    $$\\[-3.5mm]
    where $\vec{x_i}$ stands for $x_{i,1},\ldots,x_{i,n_i}$.
\end{mydescription}
\mbox{}\\[-5mm]
\mbox{Second-order} terms are considered up the
\mbox{$\alpha$-equivalence} relation induced by stipulating that, for
every operator $\oper o$, in the term 
$\oper o\big(\ldots,\bind{\vec{x_i}}t_i,\ldots\big)$ the $\vec{x_i}$ are
bound in $t_i$.

\begin{example}
Two 
terms for the \mbox{$\lambda$-calculus}
signature~(Example~\ref{LambdaCalculusSignature}) follow:
\\[2mm]\null\hfill$
\metavar m:[1],\metavar n:[0]\sep\cdot
\vdash 
\oper{app}\big(\oper{abs}\big((x)\metavar m[\var x]\big),\metavar n[\,]\big)
\enspace ,
\quad
\metavar m:[1],\metavar n:[0]\sep\cdot
\vdash \metavar m[\metavar n[\,]] 
\enspace.
$\hfill\null\\[-4mm]
\end{example}

\vspace*{-5mm}
\subsubsection{Substitution calculus.}
The second-order nature of the syntax requires a two-level substitution
calculus~\cite{Aczel,
FioreLICS08}.  Each level respectively accounts for the substitution of
variables and metavariables, with the latter operation depending on the
former.

The operation of capture-avoiding simultaneous \emph{substitution} of terms
for variables maps 
\\[1mm]\null\hfill$
{\Mcon\sep x_1,\ldots,x_n\vdash t}
\quad \mbox{ and } \quad 
{\Mcon\sep\Con\vdash t_i\ \ (1\leq i\leq n)}
$\hfill\null\\[1mm]
to
\\[0mm]\null\hfill$
\Mcon\sep\Con\vdash t\vsubst{\subst{t_i}{x_i}}_{1\leq i\leq n}
$\hfill\null\\[2mm]
according to the following inductive definition:
\begin{itemize}
\item
  $\var x_j\vsubst{\subst{t_i}{x_i}}_{1\leq i\leq n}
  = t_j$\\[-2mm]

\item
  ${\big(\metavar m[\ldots,s,\ldots]\big)
   \vsubst{\subst{t_i}{x_i}}_{1\leq i\leq n}
   =\metavar m
   \big[\ldots,s\vsubst{\subst{t_i}{x_i}}_{1\leq i\leq
   n},\ldots\big]}$\\[-2mm]

\item
  $\big(\oper o(\ldots,(y_1,\ldots,y_k)s,\ldots)\big)
   \vsubst{\subst{t_i}{x_i}}_{1\leq i\leq n}\\[1mm]
   \mbox{}\quad = \
   \oper o\big(\ldots,
               (z_1,\ldots,z_k)
	         s\vsubst{\subst{t_i}{x_i},\subst{z_j}{y_j}}
		  _{1\leq i\leq n, 1\leq j\leq k},
               \ldots\big)$\\[1mm]
   with $z_j\not\in\dom(\Con)$ for all ${1\leq j\leq k}$
\end{itemize}

The operation of \emph{metasubstitution} of abstracted terms for
metavariables maps
\\[2mm]\null\hfill$
{\metavar m_1:[m_1],\ldots,\metavar m_k:[m_k]
  \sep\Con\vdash t}
\ \mbox{ and } \ 
{\Mcon\sep\Con,\vec{x_i}\vdash t_i
    \ \ (1\leq i\leq k)}
$\hfill\null\\[1mm]
to
\\[-1mm]\null\hfill$
\Mcon\sep\Con
  \vdash t\msubst{\metavar m_i:=(\vec{x_i})t_i}_{1\leq i\leq k}
$\hfill\null\\[2mm]
according to the following inductive definition:\\[-4mm]
\begin{itemize}
\item
  $\var x\msubst{\metavar m_i:=(\vec{x_i})t_i}_{1\leq i\leq k} 
   = \var x$\\[-2mm]

\item
  $\big(\metavar m_\ell[s_1,\ldots,s_m]\big)
     \msubst{\metavar m_i:=(\vec{x_i})t_i}_{1\leq i\leq k}
     \mbox{}\ = 
       t_\ell\vsubst{\subst{s'_j}{x_{i,j}}}_{1\leq j\leq m}$\\[1mm]
   where, for $1\leq j\leq m$, 
   $s'_j=s_j\msubst{\metavar m_i:=(\vec{x_i})t_i}_{1\leq i\leq
   k}$\\[-2mm]

\item
  $\big(\oper o(\ldots,(\vec x)s,\ldots)\big)
     \msubst{\metavar m_i:=(\vec{x_i})t_i}_{1\leq i\leq k}
     \mbox{}\ = \ 
     \oper o\big(\ldots,
                 (\vec x)s\msubst{\metavar m_i:=(\vec{x_i})t_i}_{1\leq i\leq k},
		 \ldots\big)$
\end{itemize}

\vspace*{-5mm}
\subsubsection{Presentations.} 
An \emph{equational presentation} is specified by a signature together with a
set of axioms over it, each of which is a pair of terms in context.  

\begin{example}
\label{BetaEtaAxioms}
The equational presentation of the \mbox{$\lambda$-calculus} extends the
signature of Example~\ref{LambdaCalculusSignature} with the following
equations.  
\\[2mm]\null\hfill$
\begin{array}{ll}
  (\beta)\enspace & 
  \metavar m:[1],\metavar n:[0]
  \sep
  \cdot
  \vdash
  \oper{app}\big(\oper{abs}(\,(x)\metavar m[\var x]\,),\metavar n[\,]\big)
  \equiv
  \metavar m\big[\metavar n[\,]\big]
\\[2mm]
  (\eta) & 
  \metavar f:[0]
  \sep
  \cdot
  \vdash 
  \oper{abs}\big(\,(x)\oper{app}(\metavar f[\,],\var x)\,\big)
  \equiv
  \metavar f[\,]
\end{array}
$\hfill\null\\[-4mm]
\end{example}

\subsubsection{Logic.}
The rules of \emph{\mbox{Second-Order} Equational Logic} are given in
Figure~\ref{Log}.  Besides the rules for axioms and equivalence, it
consists of just one additional rule stating that the operation of
metasubstitution in extended variable contexts is a congruence.  

\begin{figure}[tbl]
\begin{minipage}{\textwidth}
\hrulefill\\[-4mm]
\begin{center}\begin{tabular}{l}
(Axiom) 
\\[2mm]
\qquad
$
\begin{array}{c}
\raisebox{.5mm}{$(\Mcon\sep\Con \vdash s \equiv t)\in E$}
\\ \hline
\raisebox{-.5mm}{$\Mcon\sep\Con \vdash s \equiv t$}
\end{array}$
\\[5mm]
(Equivalence)
\\[2mm]
$
\begin{array}{c}
  \raisebox{.5mm}{$\Mcon\sep\Con\vdash t$}
  \\ \hline
  \raisebox{-.5mm}{$\Mcon\sep\Con\vdash t\equiv t$}
\end{array}
\qquad
\begin{array}{c}
  \raisebox{.5mm}{$\Mcon\sep\Con\vdash s\equiv t$}
  \\ \hline
  \raisebox{-.5mm}{$\Mcon\sep\Con\vdash t\equiv s$}
\end{array}
\qquad
\begin{array}{c}
  \raisebox{.5mm}{$\Mcon\sep\Con\vdash s\equiv t
  \qquad
  \Mcon\sep\Con\vdash t\equiv u$}
  \\ \hline
  \raisebox{-.5mm}{$\Mcon\sep\Con\vdash s\equiv u$}
\end{array}
$
\\[5.5mm]
(Extended metasubstitution)
\\[2mm]
\qquad
$
\begin{array}{c}
\raisebox{.5mm}{$
\metavar m_1:[m_1]
,\ldots,
\metavar m_k:[m_k]
\sep
\Con
\vdash 
s \equiv t
$}
\enspace\qquad
\raisebox{.5mm}{$\Mcon\sep\aCon,\vec{x_i}
\vdash s_i\equiv t_i
\quad(1\leq i\leq k)$}
\\ \hline
\raisebox{-1.5mm}{$
\Mcon \sep \Con,\aCon \vdash 
s\msubst{\metavar m_i := (\vec{x_i})s_i}_{1\leq i\leq k}
\equiv 
t\msubst{\metavar m_i := (\vec{x_i})t_i}_{1\leq i\leq k}
$}
\end{array}$
\\[-2mm]
\end{tabular}\end{center}
\caption{\emph{Second-Order Equational Logic}.}
\label{Log}
\hrulefill
\end{minipage}
\end{figure}

We note the following basic result from~\cite{FioreHur10}:
%
    \emph{\mbox{Second-Order} Equational Logic} is a conservative extension of
    \emph{(\mbox{First-Order}) Equational Logic}.


\section{Second-Order Universal Algebra}
\label{SO:UniversalAlgebra}

The model theory of Fiore and Hur~\cite{FioreHur10} for \mbox{second-order}
equational presentations is recalled.  This is presented here in concrete
elementary terms, but could have also been given in abstract monadic terms.
The reader is referred to~\cite{FioreHur10} for the latter perspective.

\vspace*{-4.5mm}
\subsubsection{Semantic universe.}
We write $\F$ for the free cocartesian category on an object.  Explicitly,
it has set of objects~$\Nat$ and morphisms~$m\rightarrow n$ given by
functions~${\card m\rightarrow\card n}$, where, for $\ell\in\Nat$,
$\card\ell\eqdef\setof{1,\ldots,\ell}$.

We will work within and over the semantic universe~$\Set^\F$ of sets in
variable contexts~\cite{FiorePlotkinTuri}.
We write $\yon$ for the Yoneda 
embedding~$\F^\op\hspace{.075mm}\rightembedding\Set^\F$.

\subsubsection{Substitution.}
We recall the \emph{substitution monoidal structure} in 
semantic universes~\cite{FiorePlotkinTuri}.
It has tensor unit and tensor product respectively given by $\yon1$ and
$X\stensor Y \eqdef \coend^{k\in\F} X(k) \times Y^k$.

A monoid~$\yon1\hspace{-.1mm}\midstackrightarrow{\mbox{\small$\nu$}} 
A\hspace{-.1mm}\midstackleftarrow{\mbox{\small$\varsigma$}} A\stensor A$
for the substitution monoidal structure equips $A$ with substitution
structure.  In particular, the
map~${\nu_k 
\eqdef 
  ( \yon k\iso(\yon 1)^k
     \midstackrightarrow{\raisebox{1.5mm}{\small$\nu^k$}}
     A^k )}$ 
induces the embedding
\\[2mm]\null\hfill{$ 
\big(A^{\syon n}\times A^n\big)(k)
\rightarrow
A(k+n)
\times 
A^k(k)
\times 
A^n(k)
\rightarrow
\big(A\stensor A\big)(k) 
$}\hfill\null\\[2mm] 
which together with the multiplication yield a \emph{substitution operation}
\\[1.25mm]\null\hfill
$
\varsigma_n:
A^{\syon n}\times A^n \rightarrow A
\enspace.
$\hfill\null\\[1.25mm]
These substitution operations provide the interpretation of metavariables.

\vspace*{-3mm}
\subsubsection{Algebras.}
Every signature~$\Sig$ induces a \emph{signature endofunctor} on $\Set^\F$
given by
$\textstyle
\EndoFun\Sig X
\eqdef
\coprod_{\oper o:(n_1,\ldots,n_k)\,\mbox{\scriptsize in}\,\Sigma} 
  \prod_{1\leq i\leq k} X^{\syon n_i}$. 
\mbox{$\EndoFun\Sig$-algebras}~${\EndoFun\Sig X\rightarrow X}$ provide an
interpretation~${\sem{\oper o}_X:\prod_{1\leq i\leq k} X^{\syon
n_i}\rightarrow X}$ for every operator~${\oper o:\tup{n_1,\ldots,n_k}}$ in
$\Sig$.

We note that there are canonical natural isomorphisms\\[-5mm]
\begin{eqnarray*}\label{CanonicalIso1}
\textstyle
\coprod_{i\in I} (X_i\stensor Y) 
& \iso &\textstyle
\big(\coprod_{i\in I} X_i\big)\stensor Y 
\\[1.5mm]\label{CanonicalIso2}
\textstyle
\big(\prod_{1\leq i\leq n} X_i\big)\stensor Y 
& \iso &\textstyle
\prod_{1\leq i\leq n} (X_i\stensor Y) 
\end{eqnarray*}\\[-4mm]
and, for all points~$\eta:\yon1\rightarrow Y$, natural extension
maps\\[-2mm]
\begin{equation*}\label{ExtensionMap}
  \eta^{\#_n}:
X^{\,\syon n}\stensor Y
\rightarrow
(X \stensor Y)^{\syon n}
\enspace.
\end{equation*}
These constructions equip every signature endofunctor with a \emph{pointed
strength} 
$\varpi_{X,\syon1\myarrow{2}{}{} Y}
 :\EndoFun\Sig(X)\stensor Y \rightarrow \EndoFun\Sig(X\stensor Y)$. 
See~\cite{FioreLICS08} for details. 

\vspace*{-3mm}
\subsubsection{Models.}
The models that we are interested in (referred to as
\emph{\mbox{$\Sigma$-monoids}} in~\cite{FiorePlotkinTuri,FioreLICS08}) are
algebras equipped with a compatible substitution structure.
For a signature~$\Sig$, we let $\Mod\Sig$ be the category of
\emph{\mbox{$\Sigma$-models}} with objects
$A\in\Set^\F$ equipped with an \mbox{$\EndoFun\Sig$-algebra}
structure~${\alpha:\EndoFun\Sig A\rightarrow A}$
and a monoid structure\linebreak
${\yon1\midstackrightarrow{\mbox{\small$\nu$}} A
\midstackleftarrow{\mbox{\small$\varsigma$}} A\stensor A}$ 
that are compatible in the sense that the diagram\\[-3mm]
$$
\xymatrix@C=40pt@R=20pt{
\EndoFun\Sig(A)\stensor A \ar[d]_-{\mbox{\small$\alpha\stensor A$}}
\ar[r]^-{\mbox{\small$\varpi$}
_{A,\nu
}} 
& \EndoFun\Sig(A\stensor A)
\ar[r]^-{\sEndoFun\Sig\mbox{\small$\varsigma$}} & 
\EndoFun\Sig(A) \ar[d]^-{\mbox{\small$\alpha$}}
\\
A\stensor A \ar[rr]_-{\mbox{\small$\varsigma$}} & & A
}
$$\\[-1.5mm]
commutes.  Morphisms are maps that are both \mbox{$\EndoFun\Sig$-algebra}
and monoid homomorphims.

\vspace*{-3mm}
\subsubsection{Semantics.}
For $\Mcon=(\metavar m_1:[m_1] \ldots, \metavar m_k:[m_k])$ and
${\Gamma=(x_1,\ldots,x_n)}$, the interpretation of a
term~${\Mcon\sep\Gamma\vdash t}$ in a model~$A$ is a morphism
$$\textstyle
\sem{\Mcon\sep\Con\vdash t}_A: 
  \sem{\Mcon\sep\Con}_A \rightarrow A
\enspace,
$$
where
$
\textstyle
\sem{\Mcon\sep\Con}_A
= 
\prod_{1\leq i\leq k} A^{\syon m_i} \times \yon n 
$, 
given by structural induction as follows:\\[-3.75mm]
\begin{itemize}
  \item
    $\sem{\Mcon\sep\Con\vdash x_j}_A$ is the composite
$
\textstyle
\sem{\Mcon\sep\Con}_A
\stackrightarrow{\pi_2}
\yon n
\medstackrightarrow{\nu_n}
A^n
\stackrightarrow{\pi_j}
A$.\\[-2mm]

  \item
    $\sem{\Mcon\sep\Con\vdash
          \metavar m_i[t_1,\ldots,t_{m_i}]}_A$ 
    is the composite\\[-2mm]
$$
\textstyle
\sem{\Mcon\sep\Con}_A
\xymatrix@C=35pt{\ar[r]^-{\pair{\pi_i\icomp\pi_1,f}}&}
A^{\syon m_i} 
\times 
A^{m_i}
\textstyle
\medstackrightarrow{\varsigma_{m_i}}
A
$$\\[-3.5mm]
where
$f = \bigseq{ \sem{\Mcon\sep\Con\vdash t_j }}_{1\leq j\leq m_i}$.\\[-1mm]
  
  \item
    For ${\oper o:(n_1,\ldots,n_\ell)}$,\\[-3mm]
$$
    \sem{\Mcon\sep\Con\vdash
           \oper o\big((\vec{y_1})t_1,\ldots,(\vec{y_\ell})t_\ell\big)}
$$\\[-4.5mm]
    is the composite
$
\textstyle
\sem{\Mcon\sep\Con}_A
\xymatrix@C=40pt{\ar[r]^-{\seq{f_j}_{1\leq j\leq \ell}}&}
\prod_{1\leq j\leq \ell} A^{\syon n_j}
\medstackrightarrow{\sem{\oper o}_A}
A
$ 
where $f_j$ is the exponential transpose of\\[-4mm]
$$\hspace{-5.5mm}\textstyle
\prod_{1\leq i\leq k}A^{\syon m_i}
\times 
\yon n\times\yon n_j
\iso
\prod_{1\leq i\leq k}A^{\syon m_i}
\times 
\yon(n+n_j)
\xymatrix@C=60pt{\ar[r]^-{\sem{\Mcon\sep\Con,\vec{y_j}\vdash
t_j}_A}&}
A
\enspace.
$$
\end{itemize}

\vspace*{-3mm}
\subsubsection{Equational models.}
We say that a model~$A$ \emph{satisfies} $\Mcon\sep\Con\vdash s\equiv t$,
for which we use the notation ${A\models(\Mcon\sep\Con\vdash s\equiv t)}$,
iff $\sem{\Mcon\sep\Con\vdash s}_A = \sem{\Mcon\sep\Con\vdash t}_A$.

\shorten{
\begin{example}
For the signature of the \mbox{$\lambda$-calculus}
(Example~\ref{LambdaCalculusSignature}), a model 
$$\begin{array}{c}
  A^{\yon1} \medstackrightarrow{\oper{abs}} A
\enspace,\quad
A\times A\medstackrightarrow{\oper{app}} A
\\[1mm]
\yon1 \medstackrightarrow{\oper{var}} A \leftarrow A\stensor A
\end{array}
$$
satisfies the $(\beta)$ and $(\eta)$ axioms (Example~\ref{BetaEtaAxioms})
iff the following diagrams commute.
$$
\xymatrix@C=25pt{
\ar@{}[dr]|-{(\beta)}
\ar@/^1.5pc/[dr]^-{\mathsf{subst}} 
A^{\yon1}\times A 
\ar[d]_-{\oper{abs}\times\id} 
\\
A \times A \ar[r]_-{\oper{app}} & A
}
\qquad\qquad
\xymatrix@C=25pt{
\ar@{}[dr]|-{(\eta)}
A
\ar[d]_-{\lambda\mbox{\normalsize(}\oper{app}\icomp(\id\times\oper{var})\mbox{\normalsize)}}
\ar@/^1.5pc/[dr]^-{\id} 
& \\ 
A^{\yon1} \ar[r]_-{\oper{abs}} & A
}
$$
\end{example}
}

For an equational presentation~$(\Sig,E)$, we write $\Mod{(\Sig,E)}$ for the
full subcategory of $\Mod\Sig$ consisting of the \mbox{$\Sig$-models} that
satisfy the axioms $E$.  

\shorten{
\begin{remark*}
For an equational presentation~$(\Sig,E)$, with underlying set of
types~$T$, the forgetful
functor~$\Mod{(\Sig,E)}\rightarrow(\Set^{\F[T]})^T$ is monadic, the
induced monad is finitary and preserves epimorphisms, and the category of
models is complete and cocomplete.
The monadic view of \mbox{second-order} universal algebra will be
expounded upon elsewhere.
\end{remark*}
}

\vspace*{-3mm}
\subsubsection{Soundness and
completeness~\cite{FioreHur10}.}\mbox{}\\[-5mm]
\begin{quote}
    For an equational presentation~$(\Sigma,E)$, the judgement
    $\Mcon\sep\Con\vdash s\equiv t$ is derivable from $E$ iff
    $A\models(\Mcon\sep\Con\vdash s\equiv t)$ for all \mbox{$(\Sig,E)$-models}
    $A$.
\end{quote}


\section{Second-Order Algebraic Theories}
\label{SO:AlgebraicTheories}

We introduce the notion of unityped \mbox{second-order} algebraic theory and
establish it as the categorical counterpart to that of \mbox{second-order}
equational presentation.  The generalisation to the \mbox{multi-typed} case
should be evident.

\begin{remark*}
Having omitted the monadic view of \mbox{second-order} universal algebra, the
important role played by the monadic perspective in our development will not
be considered here.  
\end{remark*}

\subsubsection{Theory of equality.}
The 
theory of equality plays a pivotal role in the definition of algebraic
theory.  Thus, we proceed first to identify the \mbox{second-order}
algebraic theory of equality. 
This we do both in 
syntactic and 
semantic terms.  
The (\mbox{first-order}) algebraic theory of equality is then considered
from this new perspective.

\hide{
\begin{notation}
For $n\in\Nat$, we let $\card n=\setof{1,\ldots,n}$.
\end{notation}
}

The syntactic viewpoint leads us to 
define 
the category~$\EqTh$ with set of objects $\Nat^*$ and
morphisms~${\tup{m_1,\ldots,m_k}\rightarrow\tup{n_1,\ldots,n_\ell}}$ given by
tuples\\[-2mm]
$$\seq{\, \metavar m_1:[m_1], \ldots, \metavar m_k:[m_k]
          \sep 
	  x_1, \ldots, x_{n_i}
          \vdash 
          t_i 
       \,}_{i \in \card\ell}$$\\[-4mm]
of terms under the empty signature.
The identity on $\tup{ m_1, \dots, m_k }$ is given by\\[-2mm]
$$
\seq{\, \metavar m_1:[m_1], \dots, \metavar m_k:[m_k]
        \sep 
	x_1, \dots, x_{m_i}
        \vdash
        \metavar m_i[ x_1, \dots, x_{m_i} ] 
     \, }_{i \in \card k} 
\enspace;
$$\\[-4mm]
whilst the composition of\\[-2mm]
$$ 
\seq{\, \metavar m_1:[\ell_1], \dots, \metavar m_i:[\ell_i] 
        \sep
	x_1,\ldots,x_{m_p}
        \vdash 
        s_p 
     \,}_{p \in \card j} 
: \tup{\ell_1,\ldots,\ell_i} \rightarrow \tup{m_1,\dots,m_j}
$$\\[-4mm]
and\\[-2mm]
$$ 
\seq{\, \metavar m_1:[m_1], \dots, \metavar m_j:[m_j] 
        \sep
	x_1,\ldots,x_{n_q}
        \vdash 
        t_q 
     \,}_{q \in \card k} 
: \tup{m_1,\ldots,m_j} \rightarrow \tup{n_1,\dots,n_k}
$$\\[-4mm]
is given by metasubstitution as follows:\\[-2mm]
$$ 
\seq{\,
     \metavar m_1:[\ell_1],\ldots,\metavar m_i:[\ell_i]
     \sep
     x_1,\ldots,x_{n_q}
     \vdash 
     t_q\msubst{ \metavar m_p := (x_1,\ldots,x_{m_p})s_p }_{p\in\card j}
     \, }_{q\in\card k}
\enspace . 
$$ 

\shorten{
\begin{remark*}
Write $\eqpres E_0$ for the unityped equational presentation with no operators
and no equations, and let $\Kl_0$ be the Kleisli category induced by the
monadic forgetful functor~$\Mod{\eqpres E_0}\rightarrow\Set^{\F[1]}$.  Then,
the category~$\EqTh$ is (equivalent to) the opposite of the restriction of 
$\Kl_0$ to finite sums of representables.  
\end{remark*}
}

The category~$\EqTh$ is strict cartesian, with terminal object given by
the empty sequence 
and binary products given by concatenation.  Furthermore, the object
${\tup{0}\in\EqTh}$ is exponentiable.  Indeed, the
exponential object~$\tup{0}\Exp\tup{m_1,\ldots,m_k}$ is 
${\tup{m_1+1,\ldots,m_k+1}}$ with evaluation 
map\\[2mm]\null\hfill$
\tup{ m_1+1,\ldots,m_k+1,0}\rightarrow\tup{m_1,\ldots,m_k}
$\hfill\null\\[0mm]
given by\\[-1.5mm]
$$
\left\langle
\begin{array}{l}
\metavar m_1:[m_1+1],\ldots,\metavar m_k:[m_k+1],\metavar m_{k+1}:[0]
     \sep
     x_1,\ldots,x_{m_i}
\\[2mm]
\qquad \vdash \
     \metavar m_i\big[x_1,\ldots,x_{m_i},\metavar m_{k+1}[\,]\big]
\end{array}
\right\rangle_{i\in\card k}
$$\\[-1mm]
In fact, this structure provides a semantic characterisation of $\EqTh$.

\begin{lemma}[Universal property of $\EqTh$]
The category~$\EqTh$, together with the object~${\tup{0}\in\EqTh}$, is
initial amongst cartesian categories equipped with an exponentiable object
(with respect to cartesian functors that preserve the 
exponentiable object).
\end{lemma}
Loosely speaking, then, 
$\EqTh$ is the free (strict) cartesian category on an exponentiable
object.
%

\vspace*{-3.25mm}
\subsubsection{Algebraic theories.}
We extend Lawvere's fundamental notion of (\mbox{first-order}) algebraic
theory~\cite{LawvereThesis
} to second order.

\begin{definition}[Second-order algebraic theories]
\label{SO:AlgebraicTheoriesDefinition}
A \emph{second-order algebraic theory} consists of a 
cartesian category~$\algth T$ 
and 
a strict cartesian identity-on-objects functor~$\EqTh \rightarrow \algth T$
that preserves the exponentiable object~$\tup{0}$. 
\end{definition}
The most basic example is the \emph{\mbox{second-order} algebraic theory
of equality} given by $\EqTh$ (together with the identity functor).

Every \mbox{second-order} algebraic theory has an underlying
(\nombox{first-order}) algebraic theory.  
To formalise this, recall that the \mbox{(first-order)} algebraic theory of
equality~$\FreeCart=\F^\op$ is the free (strict) cartesian category on an
object and consider the unique cartesian functor~${\FreeCart\rightarrow\EqTh}$
mapping the generating object to the exponentiable object.  Then, the
(\mbox{first-order}) algebraic theory underlying
~${\EqTh\rightarrow\algth T}$ is 
$\FreeCart\rightarrow\algth T_0$ for
$\FreeCart\rightarrow\algth T_0\rightembedding\algth T$ the
identity-on-objects/full-and-faithful factorisation of
$\FreeCart\rightarrow\EqTh\rightarrow\algth T$.
In particular,
~$\FreeCart$ underlies
~$\EqTh$.

\vspace*{-4mm}
\subsubsection{The theory of a presentation.}
For a \mbox{second-order} equational presentation~$\eqpres E$, the
\emph{classifying category}~$\EqTh(\eqpres E)$ has set of objects~$\Nat^*$
and morphisms~$\vec m\rightarrow\vec n$, say with 
$\vec m=\tup{m_1,\ldots,m_k}$ and $\vec n=\tup{n_1,\ldots,n_\ell}$, given
by tuples\\[-2mm]
$$
\bigseq{\, \eqclass{\,
             \metavar m_1:[m_1], \ldots, \metavar m_k:[m_k]
             \sep 
              x_1, \ldots, x_{n_i}
             \vdash 
	     t_i \,}_{\eqpres E}
	     \,}_{i \in \card\ell}
$$\\[-4mm]
of equivalence classes of terms under the equivalence relation that identifies
two terms 
iff they are provably equal from $\eqpres E$ in 
\emph{Second-Order Equational Logic}.  (Identities and composition are defined
on representatives as in $\EqTh$.)
%

\shorten{
\begin{remark*}
Write $\Kl_{\eqpres E}$ for the Kleisli category induced by the monadic
forgetful functor~$\Mod{\eqpres E}\rightarrow\Set^{\F[1]}$ associated to a
unityped equational presentation~$\eqpres E$.  The 
category~$\EqTh(\eqpres E)$ is (equivalent to) the opposite of the
restriction of $\Kl_{\eqpres E}$ to finite sums of representables.  
\end{remark*}
}

\vspace*{-1mm}
\begin{lemma}
For a \mbox{second-order} equational presentation~$\eqpres E$, the
category~$\EqTh(\eqpres E)$ together with the canonical
functor~$\EqTh\rightarrow\EqTh(\eqpres E)$ is a \mbox{second-order} algebraic
theory.
\end{lemma}
\vspace*{-1mm}
We refer to $\EqTh\rightarrow\EqTh(\eqpres E)$ as the \mbox{second-order}
algebraic theory of $\eqpres E$.

\vspace*{-4mm}
\subsubsection{The presentation of a theory.}
The \emph{internal language}~$\IntEqPres(T)$ of a \mbox{second-order}
algebraic theory~$T: \EqTh \rightarrow \algth T$ is the 
\mbox{second-order} equational presentation defined as
follows:\\[-5mm]
\begin{mydescription}
\item[(Operators)]
For every $f:\tup{m_1,\ldots,m_k}\rightarrow\tup{n}$ in $\algth T$, we
have an operator $\oper o_f$ of arity 
$\tup{m_1,\ldots,m_k,\underbrace{0,\dots,0}_{\mbox{\scriptsize $n$
times}}}$.\\[1mm]

\item[(Equations)]
Setting\\[-2mm]
{
$$
\hspace{-7mm}
\oper t_f 
= 
\oper o_f
  \big( 
    (x_1,\ldots,x_{m_1})\metavar m_1[x_1,\ldots,x_{m_1}] ,
    \ldots ,
    (x_1,\ldots,x_{m_k})\metavar m_k[x_1,\ldots,x_{m_k}] ,
    x_1,\ldots,x_n
  \big)
  $$}\\[-4mm]
for every $f:\tup{m_1,\ldots,m_k}\rightarrow\tup{n}$ in $\algth T$, we
have\\[-2.25mm]
\begin{itemize}
\item
  $
  \metavar m_1:[m_1],\ldots,\metavar m_k:[m_k]
  \sep
  x_1,\ldots,x_n
  \vdash 
  s \equiv \oper t_{T\seq s}
  $\\[1mm]
  for every $\seq{s}:\tup{m_1,\ldots,m_k}\rightarrow\tup{n}$ in $\EqTh$,\\[-2mm]

\item
  $
  \metavar m_1:[m_1],\ldots,\metavar m_k:[m_k]
  \sep
  x_1,\ldots,x_n
  \vdash 
  \oper t_h 
  \equiv 
  \oper t_g
    \msubst{\metavar m_i:=(x_1,\ldots,x_{n_i})\oper t_{f_i}}_{1\leq i\leq \ell}
    $\\[1mm]
  for every $h: \tup{m_1,\dots,m_k} \rightarrow \tup{n}$, 
  $g: \tup{n_1,\dots,n_\ell} \rightarrow \tup{n}$, and 
  $f_i: \tup{m_1,\dots,m_k}\rightarrow\tup{ n_i }$, $1\leq i \leq \ell$,
  such that $h = g \comp \seq{f_1,\dots,f_\ell}$ in $\algth
  T$.
\end{itemize}
\end{mydescription}

\vspace*{-5mm}
\subsubsection{Algebraic translations.}
For \mbox{second-order} algebraic theories $T: \EqTh\rightarrow\algth T$
and $T': \EqTh\rightarrow\algth T'$, a \mbox{second-order} 
\emph{algebraic translation} $T\rightarrow T'$ is a 
functor~$F: \algth T\rightarrow\algth T'$ such that $T'=F\icomp T$.
We write~$\SOAT$ for the category of \mbox{second-order} algebraic
theories and algebraic translations.
\pagebreak

\begin{theorem}[Theory/presentation correspondence]
\label{Theory=Presentation}
Every \mbox{second-order} algebraic theory~$T:\EqTh\rightarrow\algth T$ is
isomorphic to the \mbox{second-order} algebraic theory of its associated
equational presentation $\EqTh\rightarrow\EqTh(\IntEqPres(T))$.  
\end{theorem}

\vspace*{-4.5mm}
\section{\mbox{Second-Order} Syntactic Translations}
\label{Second-OrderSyntacticTranslations}
\vspace*{-1mm}

We introduce the notion of \emph{syntactic translation} between
\mbox{second-order} equational presentations.  This we justify by
establishing its equivalence with that of algebraic translation between
the associated \mbox{second-order} algebraic theories.

\vspace*{-3mm}
\subsubsection{Signature translations.}
A \emph{syntactic translation}~$\trans: \Sig \rightarrow \Sig'$ between
\nombox{second-order} signatures is given by a mapping from the operators
of $\Sig$ to the terms of $\Sig'$ as follows:\\[-3.5mm]
$$
\mbox{$\oper o:\tup{m_1,\ldots,m_k}$
}
\quad \mapsto \quad
\mbox{$
\metavar m_1:[m_1],\ldots,\metavar m_k:[m_k]\sep\cdot
\vdash
\trans_{\oper o}$
\quad . 
}
$$\\[-4mm]
Note that the term associated to an operator has an empty variable
context and that the metavariable context is determined by the arity of
the operator.

A 
translation~$\trans: \Sig \rightarrow \Sig'$ extends to a mapping from the
terms of $\Sig$ to the terms of $\Sig'$\\[-4.5mm]
$$
\mbox{$\Mcon\sep\Con\vdash t$
}
\quad \mapsto \quad
\mbox{$\Mcon\sep\Con\vdash\trans(t)$
}
$$\\[-4.25mm]
according to the following inductive definition:\\[-5mm]
\begin{itemize}
  \item 
    $\trans(x)=x$\\[-2mm]

  \item 
    $\trans\big(\metavar m[t_1,\dots,t_m]\big)
     = 
     \metavar m\big[\trans(t_1),\dots,\trans(t_m)\big]$\\[-2mm]

 \item
   $\trans\big(\oper o\big((\vec{x_1})t_1,\ldots,(\vec{x_k})t_k\big)\big)
    =
    \trans_{\oper o}
      \msubst{
      \metavar m_i := (\vec{x_i})\trans(t_i)
        }_{1\leq i\leq k}$
\end{itemize}

\begin{lemma}[Compositionality]
The extension of a syntactic translation between \mbox{second-order}
signatures commutes with substitution and metasubstitution.
\end{lemma}

\begin{example}[Continutation Passing Style]
A formalisation of the CPS~transform for the 
\mbox{$\lambda$-calculus} as a syntactic translation due to
Plotkin~\cite{PlotkinRTA} follows.  We provide it in informal notation for
ease of readability.\\[-2mm]
$$\begin{array}{rcl}
\oper{app}:\tup{0,0}
& \mapsto & 
\metavar m:[0],
\metavar n:[0]
\sep
\cdot
\vdash 
\lambda k.\, 
  \metavar m[\,] 
  \, 
  \big(\lambda m.\, m\, (\lambda \ell.\, \metavar n[\,]\, \ell)\, k\big)
\\[2mm]
\oper{abs}:\tup{1}
& \enspace\ \mapsto\enspace\ & 
\metavar f:[1]
\sep
\cdot
\vdash
\lambda k.\,
  k \, \big(\lambda x.\, (\lambda\ell.\, \metavar f[x]\,\ell) \big)
\end{array}$$
\end{example}

\shorten{
\begin{remark*}
\mbox{Second-order} syntactic translations $\Sig\rightarrow\Sig'$ are in
one-to-one correspondence with enriched natural
transformations~$\EndoFun\Sig\Rightarrow\TermMonad{\Sig'}$, for
$\EndoFun{\Sig}$ the strong endofunctor induced by $\Sig$ and
$\TermMonad{\Sig'}$ the strong monad induced by the forgetful 
functor~$\Mod{\Sig'}\rightarrow\Set^{\F[1]}$~(see~\cite{FioreLICS08}).  These
universally induce extensions~$\TermMonad\Sig\Rightarrow\TermMonad{\Sig'}$.
\end{remark*}
}

\vspace*{-5mm}
\subsubsection{Equational 
translations.}
A \emph{syntactic translation} between second-order equational
presentations $\trans: (\Sig,E)\rightarrow(\Sig,'E')$ is a 
translation~$\trans: \Sig \rightarrow \Sig'$ such that, for every 
axiom~$\Mcon\sep\Con\vdash s\equiv t$ in $E$, the 
judgement~${\Mcon\sep\Con\vdash\trans(s)\equiv\trans(t)}$ is derivable
from $E'$.

\begin{lemma}
The extension of a syntactic translation between \mbox{second-order}
equational presentations preserves \mbox{second-order} equational
derivability.
\end{lemma}

We write $\SOEP$ for the category of \mbox{second-order} equational
presentations and syntactic translations.  (The identity syntactic translation
maps an operator $\oper o:\tup{m_1,\ldots,m_k}$ to the 
term~$\oper o
        \big(\ldots,
	     (x_1,\ldots,x_{m_i})\metavar m_i[x_1,\ldots,x_{m_i}],
	     \ldots\big)$;  
whilst the composition of $\trans$ followed by $\trans'$ maps $\oper o$ to
$\trans'(\trans_{\oper o})$.)

\begin{theorem}[Presentation/theory correspondence]
\label{Presentation=Theory}
Every \mbox{second-order} equational presentation~$\eqpres E$ is
isomorphic to the \mbox{second-order} equational presentation of its
associated algebraic theory~$\IntEqPres(\EqTh(\eqpres E))$.
\end{theorem} 

\vspace*{-7mm}
\subsubsection{Syntactic and algebraic translations.}
%
A syntactic translation~${\trans: \eqpres E \rightarrow \eqpres E'}$
induces the algebraic 
translation~$\EqTh(\trans): 
               \EqTh(\eqpres E) \rightarrow \EqTh(\eqpres E')$, 
mapping 
$
\seq{\, [t_1]_{\eqpres E},\ldots,[t_\ell]_{\eqpres E} \,}
$
to 
$\seq{\, [\trans(t_1)]_{\eqpres E'},\ldots,[\trans(t_\ell)]_{\eqpres E'} \,}$.  
This gives a functor $\SOEP\rightarrow\SOAT$.
Conversely, an
%
algebraic translation~${F: T \rightarrow T'}$ induces the syntactic
translation~$\IntEqPres(F): \IntEqPres(T) \rightarrow \IntEqPres(T')$,
mapping an operator~$\oper o_f$, for
$f:\tup{m_1,\ldots,m_k}\rightarrow\tup{n}$ in $\algth T$, to the
term~$\oper t_{Ff}
        \bigvsubst{\subst{\metavar m_{k+1}[\,]\,}{\,x_1}
	        ,\ldots,
		\subst{\metavar m_{k+n}[\,]\,}{\,x_n}}$.
This gives a functor~$\SOAT\rightarrow\SOEP$.

\begin{theorem}
\label{SOAT-SOEP-Equivalence}
The categories $\SOAT$ and $\SOEP$ are equivalent.
\end{theorem}

\vspace*{-5mm}
\section{Second-Order Functorial Semantics}
\label{SO:FunctorialSemantics}
\vspace*{-1mm}

We extend Lawvere's functorial semantics for algebraic
theories~\cite{LawvereThesis} from first to second order.  

\vspace*{-4mm}
\subsubsection{Functorial models.} 
The category $\FunMod T{\lscat C}$ of (\mbox{set-theoretic})
\emph{functorial models} of a \mbox{second-order} algebraic 
theory~$T: \EqTh \rightarrow \algth T$ 
is the category of cartesian 
functors~$\algth T\rightarrow 
\Set$ and 
natural transformations between them.

Every \mbox{$\eqpres E$}-model~$A$, for a \mbox{second-order} equational
presentation~$\eqpres E$, provides a functorial 
model~$\EqTh(\eqpres E)\rightarrow\Set$ as follows:\\[-5mm]
\begin{itemize}
  \item 
    on objects, $\tup{m_1,\ldots,m_k}$ is mapped to 
    $\textstyle\prod_{1\leq i\leq k}A(m_i)$;\\[-2mm]

  \item
    on morphisms, 
    $\seq{\,
       [ \metavar m_1:[m_1],\ldots,\metavar m_k:[m_k]
         \sep
         x_1,\ldots,x_{n_i}
	 \vdash
	 t_j
	 ]_{\eqpres E}
     \,}_{j\in\card\ell}$
    is mapped to
    $\seq{\, (f_j)_{0}
       }_{1\leq j\leq\ell}$
       where 
       $\textstyle
        f_j
        :
        \prod_{1\leq i\leq k}A^{\syon m_i}
	\rightarrow
        A^{\syon n_j}$
       is the exponential transpose of 
       $\sem{ 
          \metavar m_1:[m_1],\ldots,\metavar m_k:[m_k]
          \sep
	  x_1,\ldots,x_{n_j}
	  \vdash
	  t_j
          }_A$.
\end{itemize}
As we proceed to show, every functorial model essentially arises in this
manner~(see Corollary~\ref{Fun=Alg_Corollary}).

\hide{
A syntactic translation~$\eqpres E\rightarrow \eqpres E'$ induces an 
algebraic translation~${\EqTh(\eqpres E)\rightarrow\EqTh(\eqpres E')}$
that, in turn, yields a functorial model of~$\EqTh\rightarrow\EqTh(\eqpres
E)$ in $\EqTh(\eqpres E')$.  Therefore, syntactic translations provide
syntactic models.
}

\vspace*{-4mm}
\subsubsection{Clones.} 
We need recall and develop some aspects of the theory of \emph{clones} from
universal algebra~(see~\eg~\cite{Cohn}).

\hide{
Recall that a \emph{clone} in a cartesian category is given by a 
family~$\{ C_n \}_{n \in \mathbb{N}}$ of objects in $\mathscr{C}$ together
with 
\emph{variable maps}~$\iota^{(n)}_i
: 1 \rightarrow C_n$~(${1 \leq i \leq n\in\Nat}$) 
and 
\emph{substitution maps}~$\varsigma_{m,n}
                          : C_m \times {C_n}^m \rightarrow C_n$~($m,n\in\Nat$) 
such that:
\begin{itemize}
\item 
  $\varsigma_{n,n} 
   \circ 
   \big(\id_{C_n} \times \seq{ \iota_1^{(n)}, \dots, \iota_n^{(n)} }\big) 
   = \pi_1
   : C_n\times 1 \rightarrow C_n$,
   for all $n\in\Nat$;\\[-2mm]

\item 
  $\varsigma_{m,n} \comp (\iota_i^{(m)} \times \id_{{C_n}^m}) 
   = 
   \pi_i \comp \pi_2
   : 1\times{C_n}^m \rightarrow C_n$, 
   for all $1\leq i \leq m\in\Nat$ and $n\in\Nat$;\\[-2mm]

\item 
  $\varsigma_{m,n}\comp(\varsigma_{\ell,m}\times\id_{{C_n}^m})
  \\ \mbox{}\quad = \
   \varsigma_{\ell,n}
   \comp
   \big( \id_{C_\ell}
         \times
	 \bigpair{ \varsigma_{m,n}\comp(\pi_i\times\id_{{C_n}^m})
	           }_{1\leq i\leq\ell}
   \big)
   : C_\ell \times{C_m}^\ell\times{C_n}^m\rightarrow C_n$,\\
   for all $\ell, m,n\in\Nat$.
\end{itemize}
}

\medskip
Let $C$ be an exponentiable object in a cartesian category~$\lscat C$.
Recall that the 
family~$\opclone C = \setof{ C^n\ShortRightarrow C}_{n\in\Nat}$ has a
canonical clone structure 
\\[2mm]\null\hfill$
\iota^{(n)}_i: 1\rightarrow\opclone C_n
\ (1\leq i\leq n\in\Nat)
\enspace , \quad
\varsigma_{m,n}: \opclone C_m\times{\opclone C_n}^m\rightarrow\opclone C_n
\ (m,n\in\Nat)
$\hfill\null\\[2mm]
known as the \emph{clone of operations} on $C$.  Thus, as it is the case
with every clone, the family~$\opclone C$ canonically extends to a
functor~$\F\rightarrow\lscat C: n \mapsto \opclone C_n$.

For every $m_1,\ldots,m_k\in\Nat$~(for $k\in\Nat$), $n\in\Nat$, and
$\textstyle 
f: \prod_{1\leq i\leq k} \opclone C_{m_i}\rightarrow\opclone C_n$ in
$\lscat C$ let $\wt f = \setof{ \wt f_\ell }_{\ell\in\Nat}$ be given by 
setting
\\[1mm]\null\hfill${\textstyle
\wt f_\ell 
\, = 
\Big(
\prod_{1\leq i\leq k} \opclone C_{\ell+m_i}
\iso\,
C^\ell\ShortRightarrow \prod_{1\leq i\leq k} \opclone C_{m_i}
\xymatrix@C=35pt{\ar[r]^-{C^\ell\ShortRightarrow f}&}
C^\ell\ShortRightarrow \opclone C_{n}
\iso
\opclone C_{\ell+n}
\Big)
\hspace{.075mm} .
}$\hfill\null\\[0mm]
The family~$\wt f$ is a natural 
transformation~$\prod_{1\leq i\leq k} \opclone C_{(-)+m_i} 
  \rightarrow \opclone C_{(-)+n}$
and commutes with the clone structure.  The latter in the sense that, for
\\[2mm]\null\hfill$
w_\ell
=
\Big(\,
{\opclone C_q}^p
\iso
{\opclone C_q}^p\times 1
\xymatrix@C=70pt
{\ar[r]^-{{\opclone C_\jmath}^p\times\pair{\iota^{(q+\ell)}_{q+i}}_{1\leq i\leq \ell}}&}
{\opclone C_{q+\ell}}^p\times 
{\opclone C_{q+\ell}}^\ell
\iso
{\opclone C_{q+\ell}}^{p+\ell} 
\,\Big)
$\hfill\null\\[2mm]
where $\jmath$ denotes the inclusion $\card{q}\hookrightarrow\card{q+\ell}$,
the diagram
\\[2mm]\null\hfill
$\hspace{-1mm}
\xymatrix@R=20pt@C=10pt{
& 
\prod_{1\leq i\leq k} 
  \opclone C_{p+m_i} 
  \times
  {\opclone C_{q+m_i}}^{p+m_i}
  \ar[dr]^(.55){\hspace{10mm}\prod_{1\leq i\leq k} \varsigma_{p+m_i,q+m_i}}
& 
\\
\prod_{1\leq i\leq k} \opclone C_{p+m_i} 
\times 
{\opclone C_q}^p
\ar[d]_-{\wt f_p\times w_n}
\ar[ur]^(.45){\hspace{-10mm}\pair{ \id\times w_{m_i} }_{1\leq i\leq k}}
& 
& 
\prod_{1\leq i\leq k} \opclone C_{q+m_i}
\ar[d]^-{\wt f_q}
\\
\opclone C_{p+n} \times {\opclone C_{q+n}}^{p+n}
\ar[rr]_-{\varsigma_{p+n,q+n}} & &
\opclone C_{q+n}
}$\hfill\null\\[1.5mm]
commutes for all $p,q\in\Nat$.

\smallskip
Let $\Sig$ be a 
\mbox{second-order} signature, and consider a
functorial model $S: \EqTh(\Sig)\rightarrow\Set$.  Then, the image under
the cartesian functor~$S$ of the clone of operations induced by the
exponentiable object~$\tup{0}\in\EqTh(\Sig)$ together with the
family~$\setof{
  \wt{f_{\oper o}}
  }_{\oper o:\tup{m_1,\ldots,m_k} \,\mbox{\scriptsize in}\, \Sig}$, 
where 
$f_{\oper o} 
 = \seq{\, 
   \oper o
      (\ldots,(x_1,\ldots,x_{m_i})\metavar m_i[x_1,\ldots,x_{m_i}],\ldots) 
   \,}$,
yields a \mbox{$\Sig$-model}~$\ul S\in\Set^{\F}$.

Furthermore, for all 
$f 
 = 
 \seq{\, 
   \metavar m_1:[m_1],\ldots,\metavar m_k:[m_k]\sep x_1,\ldots,x_n 
   \vdash t
 \,}$
in $\EqTh(\Sig)$ we have that the image of $\wt f$ under 
$S:\EqTh(\Sig)\rightarrow\Set$ amounts to the interpretation of $t$ in
$\ul S$.  Thus, for all \mbox{second-order} equational
presentations~$\eqpres E= (\Sig,E)$,  the \mbox{$\Sig$-model} induced by
the restriction of a functorial model~${\EqTh(\eqpres E)\rightarrow \Set}$
to $\EqTh(\Sig)$ is an \mbox{$\eqpres E$-model}.

\smallskip
The above constructions between functorial and algebraic models provide an
equivalence.

\vspace*{-1.5mm}
\begin{theorem}
\label{Fun=Alg_Theorem}
For every \mbox{second-order} equational presentation~$\eqpres E$, the
category of algebraic models~$\Mod{\eqpres E}$ and the category of
functorial models~$\FunMod{\EqTh(\eqpres E)}\Set$ are equivalent.
\end{theorem}

\vspace*{-3.75mm}
\begin{corollary}\label{Fun=Alg_Corollary}
For every \mbox{second-order} algebraic theory~$T$, the category of
functorial models~$\FunMod{T}\Set$ and the category of algebraic
models~$\Mod{\IntEqPres(T)}$ are equivalent.
\end{corollary}

\vspace*{-7mm}
\subsubsection{Algebraic functors.}
As in the first-order case, every algebraic 
translation~$F:T\rightarrow T'$ between \mbox{second-order} algebraic
theories contravariantly induces an 
\emph{algebraic functor}~${\FunMod{T'}{\lscat C}\rightarrow\FunMod{T}{\lscat C}
: S\mapsto S\icomp F}$ 
between the corresponding categories of models.  We also have the
following fundamental result.

\vspace*{-1mm}
\begin{theorem}\label{AlgebraicFunctorsThm}
The algebraic 
functor~${\FunMod{T'}{\mbox{\scriptsize$\Set$}}
          \rightarrow
	  \FunMod{T}{\mbox{\scriptsize$\Set$}}}$ 
induced by a \nombox{second-order} algebraic translation~$T\rightarrow T'$ has
a left adjoint.
\end{theorem}

\vspace*{-6mm}
\section{Concluding Remarks}
\label{SO:ConcludingRemarks}
\vspace*{-1.5mm}

\hide{
\emph{\mbox{Second-Order} Universal Algebra}, introduced
in~\cite{FioreHur10}, provides algebraic foundations for
\mbox{multi-typed} \mbox{higher-order} equational theories as specified
by \mbox{second-order} equational presentations~(in the form of schematic
identities between terms with both variables and metavariables over
signatures of \mbox{variable-binding} operators).
}

We have introduced \emph{\mbox{second-order} algebraic
theories}~%
(Section~\ref{SO:AlgebraicTheories}):
%
$(i)$~showing them to be the \mbox{presentation-independent} categorical
syntax of \mbox{second-order} equational
presentations~(Theorems~\ref{Theory=Presentation},~\ref{Presentation=Theory},
and~\ref{SOAT-SOEP-Equivalence}), and
$(ii)$~establishing that their functorial semantics amounts to
\mbox{second-order} universal algebra~(Theorem~\ref{Fun=Alg_Theorem} and
Corollary~\ref{Fun=Alg_Corollary}).
In the context of~$(i)$, our development included a notion of
\mbox{second-order} syntactic
translation~(Section~\ref{Second-OrderSyntacticTranslations}), which, in the
context of~$(ii)$, contravariantly gives rise to algebraic functors between
categories of models~(Theorem~\ref{AlgebraicFunctorsThm}).

With this theory in place, one is now in a position to: 
$(a)$~consider constructions on \mbox{second-order} equational
presentations in a categorical setting, and indeed the developments for
(\mbox{first-order}) algebraic theories on limits, colimits, and tensor
product carry over to the \mbox{second-order} setting;
$(b)$~investigate 
{conservative-extension} 
results for \mbox{second-order} equational presentations in a mathematical
framework; and 
$(c)$~study Morita equivalence for \mbox{second-order} algebraic theories.

\vspace*{-4.55mm}

\end{document}